# A Study on Herd Behavior Using Sentiment Analysis in Online Social Network

Suchandra Dutta[1], Dhrubasish Sarkar[2], Sohom Roy[3], Dipak K. Kole[4], Premananda Jana[5]
[1,2]Amity Institute of Information Technology, Amity University Kolkata, India
[3]IBM India Private Limited, Kolkata, India
[4]Dept of CSE, Jalpaiguri Government Engineering College, Jalpaiguri, India
[5]Netaji Subhas Open University, Kalyani, India
Email: [1]duttasuchandra214@gmail.com, [2]dhrubasish@inbox.com, [3]Sohom.1988@gmail.com,
[4]dipak.kole@cse.jgec.ac.in, [5]prema_jana@yahoo.com

*Abstract*— Social media platforms are thriving nowadays, so a huge volume of data is produced. As it includes brief and clear statements, millions of people post their thoughts on microblogging sites every day. This paper represents and analyze the capacity of diverse strategies to volumetric, delicate, and social networks to predict critical opinions from online social networking sites. In the exploration of certain searching for relevant, the thoughts of people play a crucial role. Social media becomes a good outlet since the last decades to share the opinions globally. Sentiment analysis as well as opinion mining is a tool that is used to extract the opinions or thoughts of the common public. An occurrence in one place, be it economic, political, or social, may trigger large-scale chain public reaction across many other sites in an increasingly interconnected world. This study demonstrates the evaluation of sentiment analysis techniques using social media contents and creating the association between subjectivity with herd behavior and clustering coefficient as well as tries to predict the election result (2021 election in West Bengal). This is an implementation of sentiment analysis targeted at estimating the results of an upcoming election by assessing the public's opinion across social media. This paper also has a short discussion section on the usefulness of the idea in other fields.

*Keywords— Sentiment Analysis, Herd behavior, Subjectivity, Polarity, Social Network Analysis, Clustering Coefficient*

## I. INTRODUCTION

In the current era, social networking sites are regularly used to share perspectives or opinions on any product, issue, case, or any breaking news from anywhere at any moment. Analysis of sentiment or opinion mining is an area that primarily revolves around the assessment of certain views.

Online Social Network (OSN) represent a formed surrounding where users share emotions and opinions daily. They have therefore become an important requirement of sentiment/opinion-related Big Data. The purpose of Sentiment Analysis (SA) is to extract feelings or opinions from texts readily accessible by various data sources, such as OSNs [1]. A substantial number of people have expanded the use of Twitter. In an extraordinary advancement in the production of user derived content. In different fields of study, including consumer opinion and behaviour analysis during general elections, this data is now being studied. Researchers are now very much interested to get them involved in the use of different social networking sites for political activities. The study of sentiment analysis is a subset of a process which is Natural Language Processing (NLP) that focuses on determining the polarity and subjectivity of textual data. Subjectivity analysis of the textual data, which will allow to evaluate the platform's efficacy in plotting tweet subjectivity for visualization on a chart. Analysis of position data polarity, measurement of site precision in plotting mood, and views of Twitter users on a map. It is known that there are almost a million sites for microblogging. Microblogging websites are simply social media networks whereby users share short, frequent updates. Users could read and write 148-character tweets on Twitter, which is very popular microblogging platforms. Messages from Twitter are often referred to as Tweets. These tweets will be used as raw data.

*Polarity and Subjectivity analysis:*

Subjectivity analysis of the text is part of the analysis of emotion, where analysts use Natural Language Processing (NLP) to classify a text as either self-centred or not self-centred. Indicators of a subjective view are the inclusion of such words, such as adjectives, adverbs, and certain classes of verbs and nouns [2]. Therefore, subjectivity research categorizes sentences as either personal beliefs or objective facts and then use subjectivity analysis to differentiate sentences that are contextual and those that are objective for a group of messaging in a survey [3]. Word characteristics are used as a measure of a statement's subjectivity, such as the use of adjectives and nouns [4]. For Natural Language Processing (NLP), Python TextBlob has been used to conduct a subjectivity analysis of the database. Python TextBlob is a popular open source NLP library for evaluating text subjectivity. The subjectivity of each tweet was calculated using the TextBlob library, which has a built-in model for calculating subjectivity. The subjectivity range runs from 0 to 1. A value close to 0 means that the value is an Objective text implies a strongly biased text with a meaning like 1. Then the decision was taken whether an expression is perceived by subjectivity analysis. Also proceeding with polarity analysis on the document to see whether it represents a positive or pessimistic viewpoint. The aim of polarity analysis is to find the text author's emotional attitude toward the topic at hand [3]. Twitter is becoming more popular as a networking tool, there have also been improved requirements in evaluating its data to assess public sentiment or opinion on a subject or a prominent individual[5]. Analysis of polarity helps to quantify text sentiment, to obtain that perspective, the requirement appears to identify the text as negative, positive, or neutral. Several instruments are perfect for small text sentiment analysis [6]. In the following work to study the polarity on collected data, Python TextBlob has been used for subjectivity/polarity analysis. Inspection of polarity using Python TextBlob, Dictionary terms for the negative and positive polarity, ratings are assigned. These polarity ratings differ from -1 to 1, where -1 is an incredibly negative sentiment, and 1 is an incredibly positive sentiment. A 0-polarity score shows a neutral sentiment.

## II. CLUSTERING COEFFICIENT

Clustering is essentially a clustering coefficient that calculates the degree to which nodes appear to cluster or join accurately. A cluster is a group of nodes that are more connected to nodes in same group than to any other nodes, and for multiple reasons, clusters may be very important, because, for example, different clusters can be a bottleneck for knowledge or a diffusion mechanism. A cluster (Eq. 1) will hold stuff in or keep stuff out.

Mainly cluster is globally speaking is the proportion of the closed triads (i.e., A, B and C are interconnected with each other) and summation of total closed triads and open triads.

$$\text{Cluster} = \frac{(closed\ triads)}{(total\ closed\ triads)+(open\ triads)} \quad \text{----------- (1)}$$

So, a network with high clustering has a higher proportion of closed triads to all triads. The degree is the number of connections that any node has a degree distribution refers to that distribution. Lowly linked means the degree of a node is decreasing and highly linked determines the degree of a node is increasing.

The fundamental measure which accounts for a network's inner structure is clustering coefficient. The clustering coefficient refers to a network's local consistency and tests the possibility that two vertices are related to a mutual neighbor. In the case of undirected networks, there are: $[E_{max} = k_i(k_i-1)]/2]$ potential edges between neighbors given the vertex $n_i$ with $k_i$ neighbors [8].

The clustering coefficient $C_i$ (Eq. 2) of the vertex $n_i$ is then defined as the ratio between the neighbors of the actual number of edges $E_i$ to the maximum number as $E_{max}$,

$$C_i = 2E_i / k_i(k_i-1) \quad \text{----------------- (2)}$$

The coefficient of clustering $C_i$ is not a function of the vertex $n_i$ itself, but its *neighbors*. The mean cluster coefficient of all vertices is the global or mean clustering coefficient $C_i = <C_i>$ of the network [5]. Highly linked vertices show a low clustering coefficient, i.e., highly linked vertices selectively link to vertices that are not mutually linked, suggesting a hierarchical structure. The clustering coefficient $C_i$ of each vertex versus degree $k_i$ (Fig-1). Does the out-degrees $k_i^{out}$ align the vertices with the in-degrees $k_i^{in}$ of their neighbors? -The degrees should also be substituted by their weighted equivalents in the case of weighted networks [7].

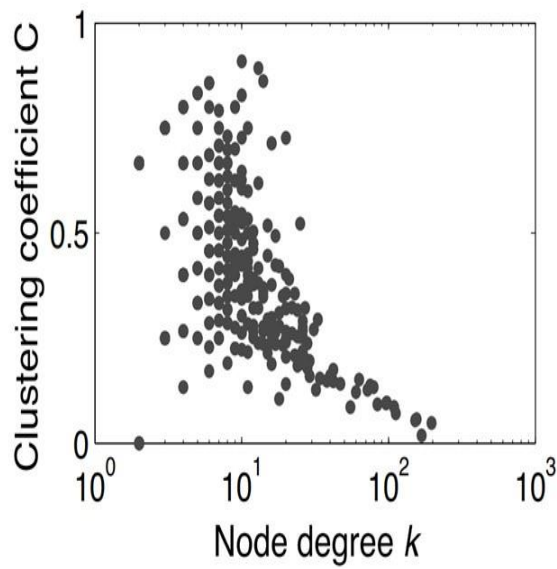

Figure 1: Graph of clustering coefficient vs node degree

## III. HERD BEHAVIOUR

An occurrence in one place, be it economic, political, or social, will trigger largescale chain responses over many other locations in an increasingly interconnected world. Herd behavior is the behavior of people without overarching direction in a group working together.

Until relatively recent times, without much reciprocal contact, such massive phenomena were investigated quite sporadically in separate social science disciplines [8]. Yet, these mass anomalies have turned out to be a topic of significant multidisciplinary concerns with developments in technologies and emerging scientific contexts.

Herding refers to an equilibrium of individuals' thoughts or actions in a group. Most notably, such fusion mostly occurs from local connections between individuals instead of by any deliberate arrangement by a central power or a leading community person. The concepts of herd behavior and collective behavior in their projected work for defining the behavior adoption on online social networks. Many researchers apply transfer learning to the field of the sentiment analysis. The study of sufficient settings for click-and view-through desires offers a complex viewpoint on social media consumer activity and social impact, evaluating social relation power in a herd behavior context [9].

## IV. RELATED WORK

Sentiment analysis from user data is becoming like a hot cake nowadays, as this information have several implications in modern world [10]. Researchers already started to analyze data from different community sites and Twitter is one of the areas where the researchers are showing more interests [11]. Product review is one of the important aspects for sentiment analysis and this will help the companies to make future marketing strategies [12]. As a subset of mining process which can be considered as opinion mining is the arear of interest for the present day's researchers [13]. Researchers are processing Twitter data as part of Big data processing [14]. Researchers also investigate and analyze the effect of share price based on Twitter tweets as part of Big data analysis [15].

In this current paper, Twitter data will be used to analyze and predict people's view towards different political groups, and it can predict the election result. And the analysis has been carried out based on current political scenarios. This work will add a new proposition in the universe of sentiment analysis.

## V. PROPOSED MODEL

This paper explores how sentiment analysis methods were used to forecast the outcome of the West Bengal election using social media content (Twitter). The searching hashtags are #BengalElection2021 and #WestBengal. With the help python programming language, Subjectivity and Polarity of the tweets can be calculated, which can help this paper to conclude.

The process and its components of the model can be explained using the below flow (Fig- 2).

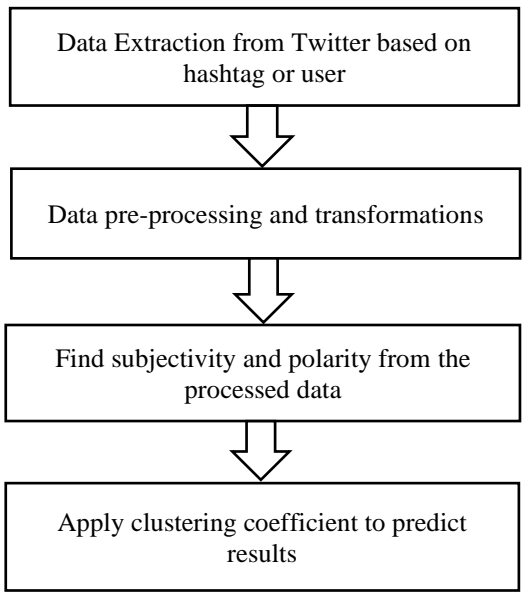

Figure 2 - Process flow

The process starts with the extraction of data from Twitter and then collected records have been pre-processed and transformed. This process is the direct application of Natural Language Processing. From the processed data subjectivity and polarity has been calculated. And finally, clustering coefficient applied to predict the result which will conclude the analysis. All the codes have been implemented using Python language and different Python libraries such as tweepy, pandas, preprocessor, nltk etc have been used.

The main focus of this paper is as follows: the subjectivity lies between 0 to 1 and the polarity lies between -1 to +1, i.e. the more numbers (tends to 1) of subjectivity refers that mostly the tweets in public opinion and the less number (tends to 0) of subjectivity refers that the tweets are factual information similarly when the polarity tends to -1 that means the polarity is negative and when the polarity tends to 1 is meant to be the polarity is positive and if the polarity of tweets is 0 that meant to be the polarity is neutral.

So, if the clustering coefficient of a high number of subjectivities is high that means the most tweets are in the public opinion, most of the people following this and most of the public accompanied the tweets as well as the particular information, as because of the most of the people are following and agree with the particular information so in reference to the herd behavior, it conveys to an equilibrium of individuals' thoughts or actions in a group. Most notably, such fusion mostly occurs from local connections between individuals instead of by any deliberate arrangement by a central power or a leading community person. So, it can be said that the tweets which have the high number of clustering coefficient of the high number of subjectivities is followed the herd behavior. That means if the opinion, as well as the thoughts of a greater number of people or group of people, is almost same then most of the people follow that opinion.

*Result for #BengalElection2021*

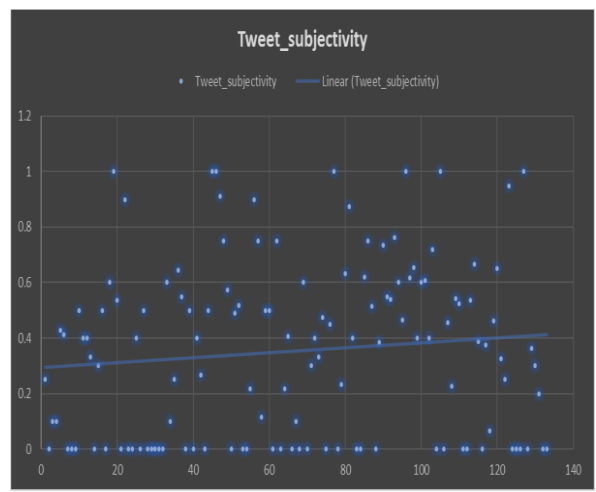

Figure 3- Subjectivity of tweets of #BengalElection2021

After fetching the tweets (134 number of tweets have been fetched) for #BengalElection2021 in the python program, getting a 17.91 % of negative tweets and 36.56% of positive tweets, and 45.52% of neutral tweets, and also detected that the subjectivity (Fig-3) of most of the tweets are having values between 0.5-0.8, which indicate the tweets in people's choice and not a piece of factual statistics and the polarity (Fig-4) of the most tweets having values between 0-1, that indicates the most of the tweets are neutral and positive. As because the clustering coefficient of the higher value of subjectivity is more, so most of the people give the similar kind of opinion and according to the herd behaviour majority of the people follow that opinion.

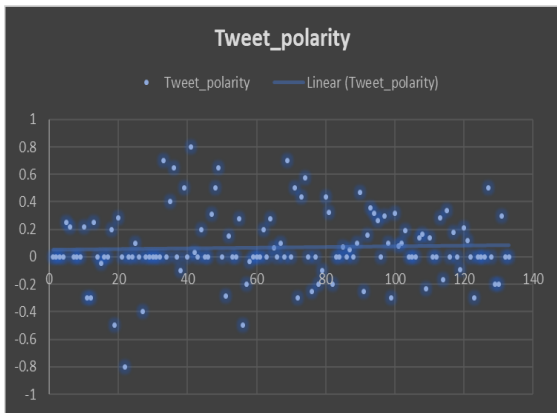

Figure 4 - Polarity of tweets of tweets of #BengalElection2021

*Result of #WestBengal*

After fetching the thousands (fetched 1933 number of tweets) of tweets for #WestBengal with the help of the python program, getting a 13.53 % of negative tweets and 38.18% of positive tweets, and 48.14% of neutral tweets, and the polarity (Fig-5) of the most of tweets range between 0-1 that indicates the most of neutral and positive tweets and observed that the subjectivity (Fig-6) of most of the tweets range between 0.5-0.8, which indicate the tweets in public opinion and not factual information. As because the clustering coefficient of a higher value of subjectivity is more so most of the people give the similar kind of opinion and in term of the herd behaviour, majority of the people follow that opinion.

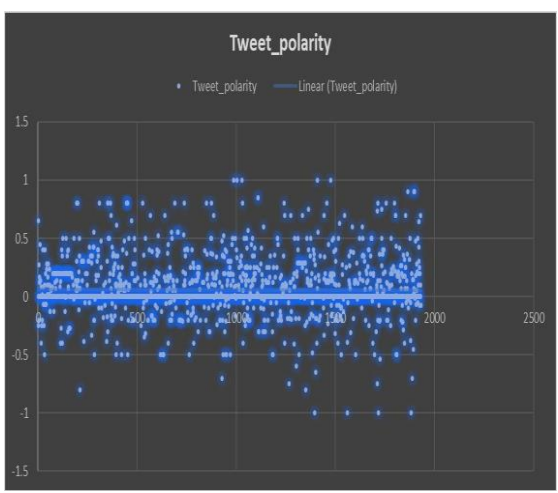

Figure 5: Polarity of the tweets of #WestBengal

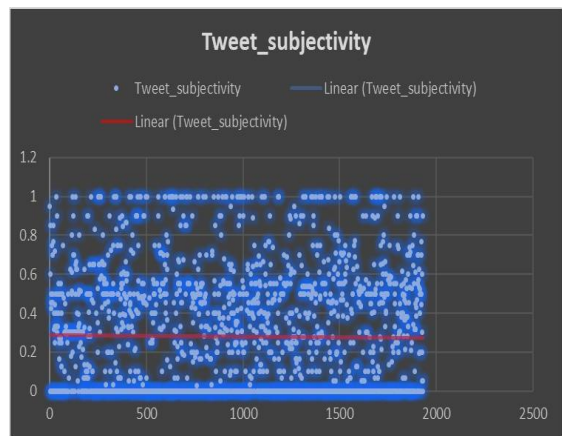

Figure 6: Subjectivity of tweets of #WestBengal

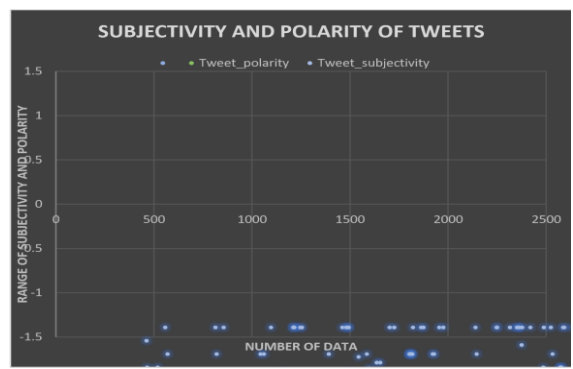

Figure 7 - Subjectivity and polarity both shown in one graph (#WestBengal)

A comparison between subjectivity and polarity (Fig-7), can be shown where subjectivity and Polarity both for the hashtag of west Bengal (#WestBengal) and they are related to the 2500 tweets which have been considered, depending on the extracted datasets.

## VI. RESULT ANALYSIS

This survey represents the outcomes of analysing sentiments as well as the herd behaviour and clustering coefficient to predict election results of West Bengal (2021) through social media content (Twitter). Therefore, the sentiment of most of the tweets are neutral and positive and the less number of the tweets are negative, as well as some of the tweets have more number subjectivity which refers that mostly it in people's choice and not a piece of factual information. So, the larger number of neutral tweets cannot conclude or predict any solution that which political party can win the 2021 election in West Bengal. Also, here the prediction can be done that the winning probability of party X is a bit higher than Y (Actual party names are kept hidden here and they are represented as X & Y). Since the higher value of subjectivity has a higher clustering coefficient, more people or users will follow this opinion according to the herd behaviour principle.

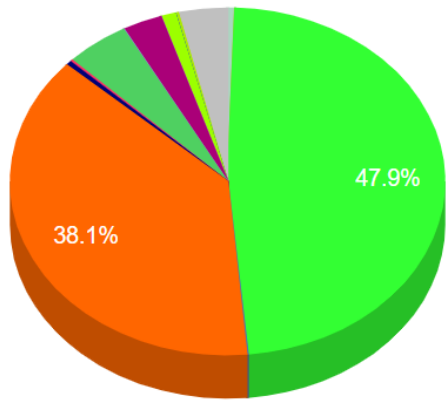

Figure 8: Published Result of West Bengal Assembly Election 2021

Now the actual result of West Bengal Assembly Election 2021 (Fig- 8) shows that party X has got 47.9% of the total vote share and party Y has got the 38.1% of the total vote share. And the model predicted the same where it claims that the probability of wining for party X is a bit higher than party Y [16]. The analysis has been carried out based on two different trending hashtags and tweets have been extracted during first week of February 2021. There are different phases of the experiment and all phases are essential to reach the result. The process, finding out the result involves the process of ETL (Extract, Transform, Load) in the first phase, and in the final phase analysis related to the clustering coefficient applied on the cleansed and transformed data. The ETL process has been implemented using Python and the records which has been extracted – are the complete tweets along with the username, hashtag used, tweet time, number of followers, retweets etc. In the next phase RAW tweets have been pre-processed and transformed so that token can formed for the further analysis. This step also takes care of the language or simply it can be referred as the implication of Natural Language Processing. These will transform the data/tweets in the form of tokens and different stop words, punctuation has been removed along with the process of lemmatization. In the final phase clustering coefficient applied to predict the outcome on the cleansed and transformed dataset and the result of this analysis has been compared and evaluated with the actual result which is the result of West Bengal Assembly Election 2021. The model shows a promising output when it is compared with the actual result of West Bengal Assembly Election 2021. There are too many predictions from other sources who claimed the completely opposite result of this model. But from twitter sentiment analysis perspective it can show the support or emotion was higher with the party X. Though the records have been extracted during first week of February 2021 and the election process took place in April 2021, and there could be several possible factors or incidents might happen in 2-3 months which could change the election result completely. But from the current scenario it can be shown that wining party was able to go with the pulse of the people of West Bengal and they are able to understand the emotion of the people in West Bengal and the same has been reflected in different tweets. Not only this, there are few contents in multiple tweets which could influence the other people and they started to change their mind. These kinds of tweets created too many followers and resulted the herd behaviours where there are certain timestamps till when people follow or supported the tweets and when the followers would be publishing/post any tweets they would be highly influenced by the previously published tweets.. In different timestamps, herd behaviours can be adopted over the Social Networks [17]. So, the model is useful and can show a new way where online platforms especially social media will take a crucial role to analyse and predict election result.

Also, the major problem is the consistency of the election predicting model. The primary purpose of any prediction method is a prediction with a high degree of precision. Because the records collected in RAW format from multiple sources include errors that may lead to erroneous predictions. More research is required to overcome the shortcomings of basic sentiment analysis systems, such as machine learning techniques.

### VII. USAGE IN OTHER SECTORS

The proposed model can be used in other sectors also; herd behaviour analysis using Facebook posts [17], crime data analysis from victims viewpoint [18], public reaction towards controlling obesity, Rational conformity and information cascades and decision making for any product marketing etc.

### VIII. CONCLUSION

Predicting election result is a little bit tricky concept, and it has been tried earlier also by researchers and different technics have been used. Those prediction process involved - from manual process to complex mathematical models. Many aspects around the society have changed because of technical advances, and people will now predict, analyse, and measure an extensive variation of critical and significant events. Due to the vast volume of data accessible from various sources, complicated actions in a minimal timeframe. Sentiment analysis of people's views from Online Social Networks like Twitter has resulted in many research conveniences. In this research paper, the prediction of West Bengal election 2021 is given with the help of sentiment analysis and clustering coefficient, and herd behaviour model.

REFERENCES

1. C. Zucco, B. Calabrese, G. Agapito, P. H. Guzzi, M. Cannataro, "Sentiment analysis for mining texts and social networks data: Methods and tools," Wiley Periodicals, https://doi.org/10.1002/widm.1333 , Vol 10(3), pp. 1-42 August 2019
2. A. Tumasjan, T. O. Sprenger, P. J. Sandner, I. M. Welpe, "Predicting elections with twitter: What 140 characters reveal about political sentiment," ICWSM 10, 1, pp. 178–185., 2010

3. U. Yaqub, N. Sharma, A. Pabrej, S. A. Chun, V. Atluri, J. Vaidya, "Analysis and Visualization of Subjectivity and Polarity of Twitter Location Data," Proceedings of the 19th Annual International Conference on Digital Goverement Reasearch: Goverenance in the Data , A67 , pp. 1-10, May 2018
4. V. A. Kharde, Prof S. Sonawane, "Sentiment analysis of Twitter data: A survey of techniques," International Journal of Computer Applications 139(11): pp.- 5-15, arXiv preprint arXiv:1601.06971 ,2016
5. N. Li, D. D. Wu, "Using text mining and sentiment analysis for online forums hotspot detection and forecast. Decision support systems," elsevier ,48, 2, pp. 354–368., 2010
6. B. O'Connor, R. Balasubramanyan, B. R. Routledge, N. A. Smith, "From tweets to polls: Linking text sentiment to public opinion time series," ICWSM 11, pp. 122-129 , 1–2.(2010)
7. B. H. Junker, F. Schreiber, "Analysis of Biological Networks, " (ISBN 978-0-470-04144-4) pp xv + 346. Hoboken: John Wiley & Sons, pp. 39, Inc. 2008.
8. T. Kameda, K. Inukai, T. Wisdom, W. Toyokawa, "The Concept of Herd Behaviour: Its Psychological and Neural Underpinnings," IOSR, pp. 61-71, December 2014.
9. J. Mattke, C. Maier, L. Reis, T. Weitzel, "Herd behavior in social media: The role of Facebook likes, strength of ties, and expertise," European Journal of Information Systems, pp. 1-25 ,2020
10. Alsaeedi, A. Khan, Mohammad, "A Study on Sentiment Analysis Techniques of Twitter Data. International Journal of Advanced Computer Science and Applications," 10. 361-374. 10.14569/IJACSA.2019.0100248, 2019
11. P. Tyagi, R. C. Tripathi, "A Review towards the Sentiment Analysis Techniques for the Analysis of Twitter Data," Proceedings of 2nd International Conference on Advanced Computing and Software Engineering (ICACSE) , Available SSRN: https://ssrn.com/abstract=3349569 or http://dx.doi.org/10.2139/ssrn.3349569, February 8, 2019
12. X. Fang, J. Zhan, "Sentiment analysis using product review data. Journal of Big Data," Journal of Big Data 2, 5 (2015). https://doi.org/10.1186/s40537-015-0015-2
13. K. Norambuena, B. Lettura, E. Villegas, Claudio. "Sentiment analysis and opinion mining applied to scientific paper reviews. Intelligent Data Analysis,". 23. 191-214. 10.3233/IDA-173807, 2019
14. B. Cornelia, P. Juergen ,"Content analysis of Twitter: Big data, big studies," In S. A. Eldridge II & B. Franklin, B. (Eds.), The Routledge Handbook to Developments in Digital Journalism Studies. Abingdon: Routledge, 2018
15. Wlodarczak. Peter, "Exploring the value of big data analysis of Twitter tweets and share prices," CSEIT1835125, 2018.
16. https://results.eci.gov.in/Result2021/partywiseresult-S25.htm?st=S25 (Accessed on 05/04/2021 02:15 PM IST)
17. D. Sarkar, S. Roy, C. Giri, D. K. Kole, "A Statistical Model to Determine the Behavior Adoption in Different Timestamps on Online Social Network. International Journal of Knowledge and Systems," IJKSS, 10(4), 1-17. doi:10.4018/IJKSS.2019100101 ,2019
18. S. Roy, S. Kundu, D. Sarkar, C. Giri., P. Jana: "Community Detection and Design of Recommendation System Based on Criminal Incidents," In Bhattacharjee D., Kole D.K., Dey N., Basu S., Plewczynski D. (eds) Proceedings of International Conference on Frontiers in Computing and Systems. Advances in Intelligent Systems and Computing, vol 1255. 71 – 80. Springer, Singapore. https://doi.org/10.1007/978-981-15-7834-2_7